\documentclass[cpp,a4paper,fleqn%
]{w-art}
\usepackage{times,cite,w-thm,amsmath,amssymb}
\theoremstyle{plain}

\theoremstyle{definition}

\usepackage[]{graphicx}
\usepackage{color}

\newcommand{\vc}{\mathbf}

\begin{document}
\DOIsuffix{theDOIsuffix}
\Volume{46}
\Month{01}
\Year{2007}
\pagespan{1}{}
\Receiveddate{XXXX}
\Reviseddate{XXXX}
\Accepteddate{XXXX}
\Dateposted{XXXX}



\title[Effective Potential Theory]{Effective Potential Theory: A Practical Way to Extend Plasma Transport Theory to Strong Coupling}


\author[S.D.\ Baalrud]{Scott D.\ Baalrud\inst{1,}%
  \footnote{Corresponding author\quad E-mail:~\textsf{scott-baalrud@uiowa.edu},
            Phone: +01\,319\,335\,1695,
            Fax: +01\,319\,335\,1753}}
\address[\inst{1}]{Department of Physics and Astronomy, University of Iowa, Iowa City, Iowa 52242, USA}
\author[K.\O.\ Rasmussen]{Kim \O.~Rasmussen \inst{2}} 
\address[\inst{2}]{Theoretical Division, Los Alamos National Laboratory, Los Alamos, NM 87545, USA}
\author[J.\ Daligault]{J\'er\^ome Daligault  \inst{2}} 
\begin{abstract}
  The effective potential theory is a physically motivated method for extending traditional plasma transport theories to stronger coupling. It is practical in the sense that it is easily incorporated within the framework of the Chapman-Enskog or Grad methods that are commonly applied in plasma physics and it is computationally efficient to evaluate. The extension is to treat binary scatterers as interacting through the potential of mean force, rather than the bare Coulomb or Debye-screened Coulomb potential. This allows for aspects of many-body correlations to be included in the transport coefficients. Recent work has shown that this method accurately extends plasma theory to orders of magnitude stronger coupling when applied to the classical one-component plasma model. The present work shows that similar accuracy is realized for the Yukawa one-component plasma model and it provides a comparison with other approaches.

 
\end{abstract}
\maketitle                   





\section{Introduction}

A significant challenge in the theory of strongly coupled Coulomb systems is to describe how many-body correlations affect transport properties. Much progress has been made using particle simulations, such as molecular dynamics (MD) and Monte Carlo, but there is still a critical need for theory that can provide insights into the underlying physical processes, as well as provide expressions for transport coefficients that can be incorporated into macroscopic (i.e., fluid) descriptions. We recently proposed the effective potential transport theory \cite{baal:13}, which is a physically motivated extension of traditional plasma theories based on the Chapman-Enskog or Grad fluid expansions. Our previous work showed that this method provides an accurate extension into the strongly coupled regime by comparing the theoretical predictions with MD simulations of the one-component plasma (OCP). The present paper shows that the theory provides a similar extension when applied to the Yukawa OCP model over a range of screening parameters ($\kappa$). 

The basic concept underlying the effective potential theory is that binary scattering events do not occur in isolation, but are instead influenced by the surrounding medium. Although the basis is a binary collision picture, a judicious choice of the effective interaction potential allows for key features of many-body correlations to be included. In some sense the effective potential concept is standard in plasma physics because screening caused by polarization is always required to avoid unphysical divergences associated with the infinite range of the bare Coulomb potential.  Although the conventional derivations introduce the truncation in the impact parameter, rather than the interaction potential, these are equivalent in the weakly coupled limit~\cite{baal:14}. Strongly coupled plasmas require account of many-body correlations in addition to screening~\cite{baus:80}. We have found that the potential of mean force contains essential features of many-body correlations that enables this approach to extend into the strong coupling regime. The potential of mean force is the potential obtained when taking two particles at fixed positions and averaging over the positions of all other particles.

This paper extends the previous tests to the Yukawa OCP model by comparing with MD simulations of self diffusion and viscosity over a range of screening parameters. It also provides a comparison with other theoretical approaches applied to positron-ion temperature relaxation.

\section{Effective Potential Theory\label{sec:ep}}

The effective potential theory is based on the Boltzmann collision operator, but where the scattering cross section is calculated from the effective interaction potential, $\phi_{ss^\prime}(r)$, between species $s$ and $s^\prime$. For application in the Chapman-Enskog or Grad near-equilibrium expansions, the transport coefficients can be expressed in terms of the oft-used $\Omega$ integrals~\cite{baal:13,baal:14}.  To facilitate the connection with weakly coupled plasma theory, the $\Omega$-integrals can  be written in the form 
\begin{equation}
\Omega_{ss^\prime}^{(l,k)} = \frac{3}{16} \frac{m_s}{m_{ss^\prime}} \frac{\nu_{ss^\prime}}{n_{s^\prime}} \frac{\Xi_{ss^\prime}^{(l,k)}}{\Xi_{ss^\prime}} , \label{eq:cints}
\end{equation}
where 
\begin{equation}
\Xi_{ss^\prime}^{(l,k)} = \frac{1}{2} \int_0^\infty d\xi\, \xi^{2k+3} e^{-\xi^2} \bar{\sigma}_{ss^\prime}^{(l)} / \sigma_o \label{eq:xilk}
\end{equation}
is a ``generalized Coulomb logarithm'' associated with the $(l,k)^\textrm{th}$ collision integral. Here
\begin{equation}
\nu_{ss^\prime} \equiv \frac{16 \sqrt{\pi} q_s^2 q_{s^\prime}^2 n_{s^\prime}}{3 m_s m_{ss^\prime} \bar{v}_{ss^\prime}^3} \Xi_{ss^\prime} \label{eq:nu}
\end{equation}
is a reference collision frequency,
\begin{equation}
\bar{\sigma}_{ss^\prime}^{(l)} = 2 \pi \int_0^\infty db\, b [1 - \cos^{l} (\pi - 2 \Theta) ]  \label{eq:sigl}
\end{equation}
is the $l^{\textrm{th}}$ momentum-transfer cross section, $\sigma_o = (\pi q_s^2 q_{s^\prime}^2)/(m_{ss^\prime}^2 \bar{v}_{ss^\prime}^4)$ is a reference cross section, $n_s$ is the number density, $q_s$ is the charge, and $m_{ss^\prime} =m_s m_{s^\prime}/(m_s + m_{s^\prime})$ is the reduced mass. Other notations are $\Xi_{ss^\prime} \equiv \Xi_{ss^\prime}^{(1,1)}$, $\xi = |\vc{v} - \vc{v}^\prime|/\bar{v}_{ss^\prime}$, $\bar{v}_{ss^\prime}^2 = v_{Ts}^2 + v_{Ts^\prime}^2$ and $v_{Ts}^2 = 2T_s/m_s$. The momentum transfer cross section depends on the scattering angle, which is computed from standard classical mechanics of two particles interacting through a central conservative potential $\phi_{ss^\prime} (r)$
\begin{equation}
\Theta = b \int_{r_o}^\infty dr\, r^{-2} \biggl[ 1 - \frac{b^2}{r^2} - \frac{2\phi_{ss'} (r)}{m_{ss^\prime} \bar{v}_{ss^\prime}^2 \xi^2} \biggr]^{-1/2} . \label{eq:theta}
\end{equation}
Here, $r_o$ is the distance of closest approach, which is determined from the largest root of the denominator in Eq.~(\ref{eq:theta}). 

The only input required to evaluate Eqs.~(\ref{eq:cints})--(\ref{eq:theta}) is the effective interaction potential $\phi_{ss^\prime}(r)$. Any central conservative potential can in principle be applied, but one that accounts for correlation effects of the background medium is required to model strongly coupled plasmas. For this we draw from equilibrium statistical mechanics and associate the effective potential with the potential of mean force. The potential of mean force is the interaction potential between two fixed particles obtained by averaging over all other particles. It is related to the pair distribution function
\begin{equation}
g_{ss^\prime}(r) = \exp [-\phi_{ss^\prime}(r)/k_B T] . \label{eq:grphi}
\end{equation}
Since this is an equilibrium approximation, the species temperatures are approximated as equal here. The task is now transferred to calculating the pair distribution function. A variety of approximations are available for this. Here we apply the hypernetted chain (HNC) approximation
\begin{subequations}
\begin{eqnarray}
g_{ss^\prime}(\vc{r})&=&\exp [- v_{ss^\prime}(\vc{r})/k_B T + h_{ss^\prime}(\vc{r}) - c_{ss^\prime}(\vc{r}) ]\label{eq:hncgr}\\
\hat{h}_{ss^\prime}(\vc{k})&=&\hat{c}_{ss^\prime}(\vc{k}) + \sum_{j=1}^N n_j \hat{h}_{sj}(\vc{k}) \hat{c}_{js^\prime} (\vc{k}) \,,  \label{eq:hnchk} 
\end{eqnarray}
\end{subequations}
where $h_{ss^\prime}(r)=g_{ss^\prime}(r)-1$ is the pair-correlation function and hats denote Fourier transformed variables.  Equations (\ref{eq:hncgr}) and (\ref{eq:hnchk}) can be efficiently solved iteratively starting from a reasonable guess for $c_{ss^\prime}({\bf r})$ [e.g., $c_{ss^\prime}({\bf r})=-v_{ss^\prime}({\bf r})/k_BT$]. In the following we explore ion transport properties using these equations where the bare potential, $v_{ss^\prime}(r)$, is modeled using the Coulomb (OCP) or screened Coulomb (Yukawa OCP) potentials.

\section{Application to the Yukawa One-Component Plasma Model}

\subsection{Self-Diffusion}

The Yukawa OCP is a reference model that treats ion motion as occurring in a bath of noninteracting neutralizing electrons~\cite{baus:80}. The ion bare potential is taken to have the form 
\begin{equation}
\frac{v(r)}{k_BT} = \frac{\Gamma}{(r/a)} e^{-\kappa r/a}  \label{eq:yukawa}
\end{equation}
where $\Gamma=q^2/ak_BT$ is the Coulomb coupling parameter,  $\kappa = a/\lambda_{sc}$ is the electronic screening parameter, and $a=(3/4\pi n)^{1/3}$ is the Wigner-Seitz radius. The classical OCP corresponds to the limit of no electron screening, $\kappa = 0$, and the Yukawa OCP to finite $\kappa$ values. Generalized Coulomb logarithms and resulting reduced transport coefficients can be characterized entirely in terms of $\Gamma$ and $\kappa$ in this model. 

\begin{figure}[b]
\sidecaption
\includegraphics[width=120mm]{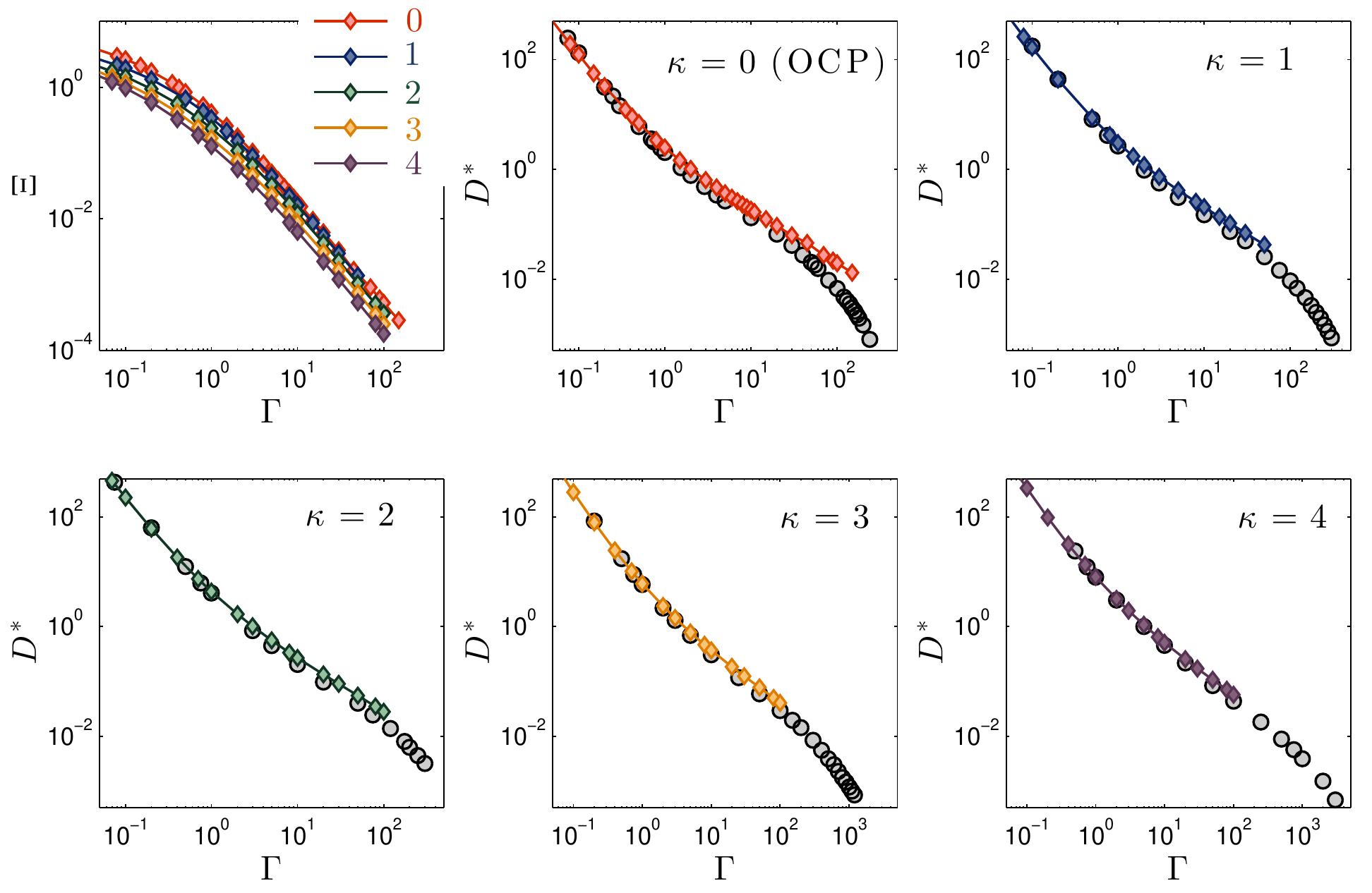}
\caption{Lowest-order generalized Coulomb logarithm, $\Xi^{(1,1)}$, and the self-diffusion coefficient of the Yukawa OCP for $\kappa = 0, 1, 2, 3$ and 4. MD simulation results are denoted by circles, and the effective potential theory results by diamonds. }
\label{fg:ocp}
\end{figure}

Figure~\ref{fg:ocp} shows the theoretical results for self-diffusion in comparison with classical MD simulations for $\kappa = 0, 1, 2, 3$ and $4$. The self-diffusion coefficient was computed from the first-order Chapman-Enskog formula~\cite{baal:14}
\begin{equation}
D^*_1 = \frac{\sqrt{\pi/3}}{\Gamma^{5/2} \Xi^{(1,1)}}.
\end{equation}
The normalization is $D^* \equiv D/(a^2 \omega_p)$ where $\omega_p=\sqrt{4\pi e^2n/m}$ is the plasma frequency. The second order correction is small over the range of coupling strengths shown here (see \cite{baal:14} for details). The effective potential was computed from the one-component version of the HNC equations using Eq.~(\ref{eq:yukawa}) as the input bare potential. The self-diffusion coefficient was extracted from MD particle data using the Green-Kubo relation; see~\cite{dali:12} for details. The figure shows that the incorporation of an effective potential that includes correlation effects enables an extension of the binary collision picture well into the strong coupling regime. The theoretical predictions eventually fail at sufficiently strong coupling. This is where strong caging effects are known to onset~\cite{donk:02,dali:12}, likely superseding the binary collision picture, even with an effective interaction potential. The effective potential theory accurately extends to higher values of $\Gamma$ at larger $\kappa$. However, the coupling strength in the Yukawa OCP is reduced by the screening parameter. For example, $\Gamma^* \simeq \Gamma \exp(-\kappa)$ is often used to approximate the coupling strength~\cite{kuzm:02} (see \cite{ott:14} for a more rigorous quantification of coupling strength).  Theoretical data is limited to $\Gamma \lesssim 100$ because the HNC numerical routine did not converge above this value. 

\subsection{Shear Viscosity}

\begin{figure}[t]
\sidecaption
\includegraphics[width=85mm]{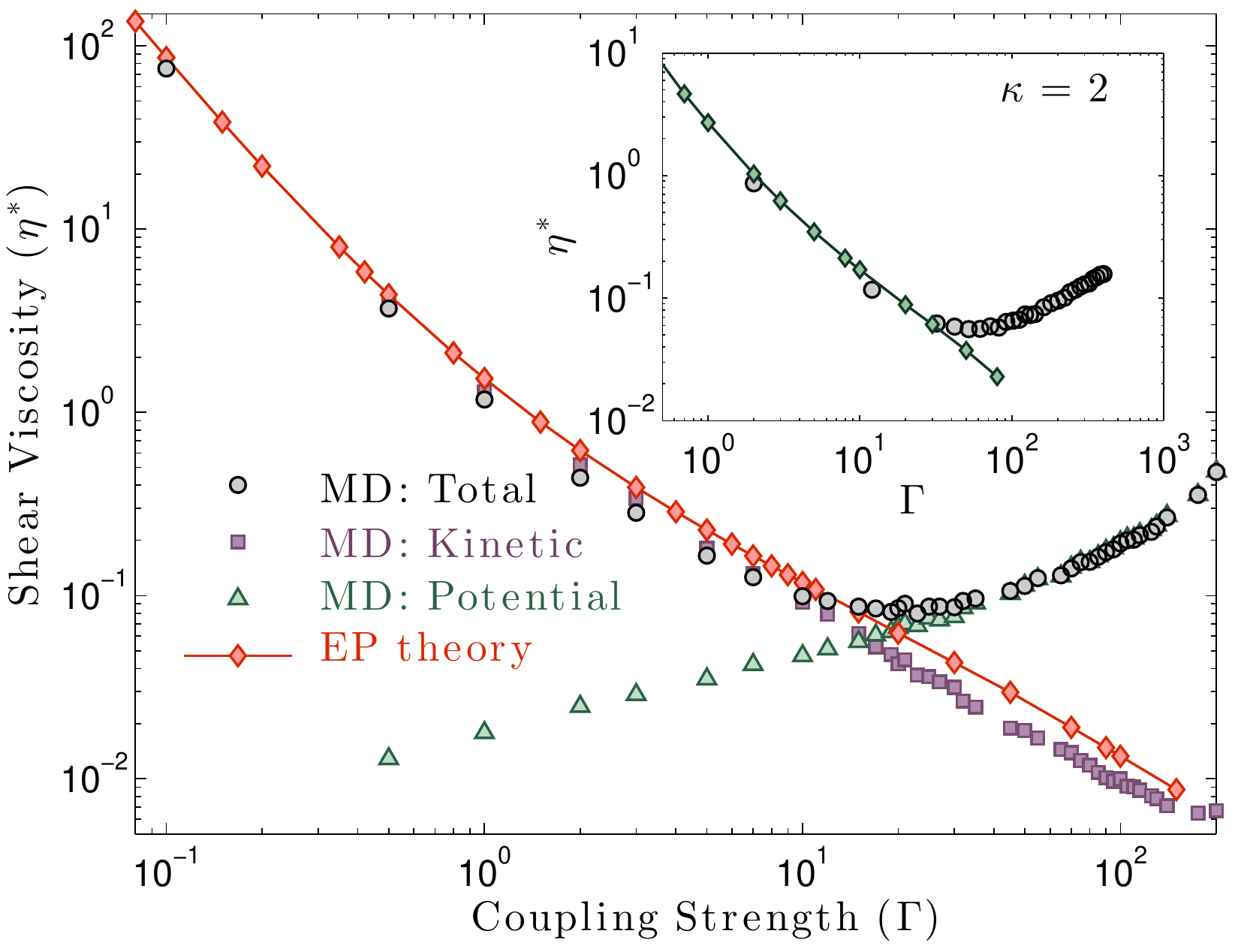}
\caption{Dimensionless shear viscosity of the Yukawa OCP for $\kappa = 0$ and 2. For $\kappa=0$, MD data is shown for the kinetic component (squares), the potential component (triangles), and the total (circles).  }
\label{fg:visc}
\end{figure}

Figure~\ref{fg:visc} shows the theoretical results for shear viscosity in comparison with classical MD simulations for $\kappa = 0$ and $2$. The viscosity coefficient was computed from the lowest-order Chapman-Enskog expression~\cite{dali:14}
\begin{equation}
\eta^*_1 = \frac{5\sqrt{\pi}}{3\sqrt{3} \Gamma^{5/2} \Xi^{(2,2)}}  .
\end{equation}
The normalization is $\eta^* \equiv \eta/(mna^2\omega_p)$. The MD simulation results where obtained using the Green-Kubo relations; see \cite{dali:14} for details. This data agrees with previous results obtained using nonequilibrium MD~\cite{donk:08}.  The comparison shows a similar range of coupling parameters over which the effective potential approach provides an accurate approximation as was found for self-diffusion. However, beyond this range the disagreement between theory and simulation is much larger for viscosity than self-diffusion. In particular, the theory does not capture the viscosity minimum. The reason for this is that although the caging aspect affects both transport processes, viscosity has additional elements that are not captured by the binary collision picture in the very strongly coupled regime. Namely, whereas diffusivity is determined entirely from particle momenta, viscosity has components from both particle momenta and electrostatic potential. Since the binary collision picture only considers the particle momenta, it captures only the kinetic contribution to viscosity. This notion is corroborated by the data in figure \ref{fg:visc}, which separates the kinetic and potential contributions (cross terms are also present, but are a negligible correction, see \cite{dali:14} for details). The effective potential theory accurately models the kinetic contribution over the entire range plotted, but the potential contributions dominate the total viscosity for $\Gamma \gtrsim 15$ (for $\kappa = 0$). The inset plot shows similar results obtained for $\kappa = 2$.

\section{Temperature Relaxation}

\begin{figure}[t]
\sidecaption
\includegraphics[width=86mm]{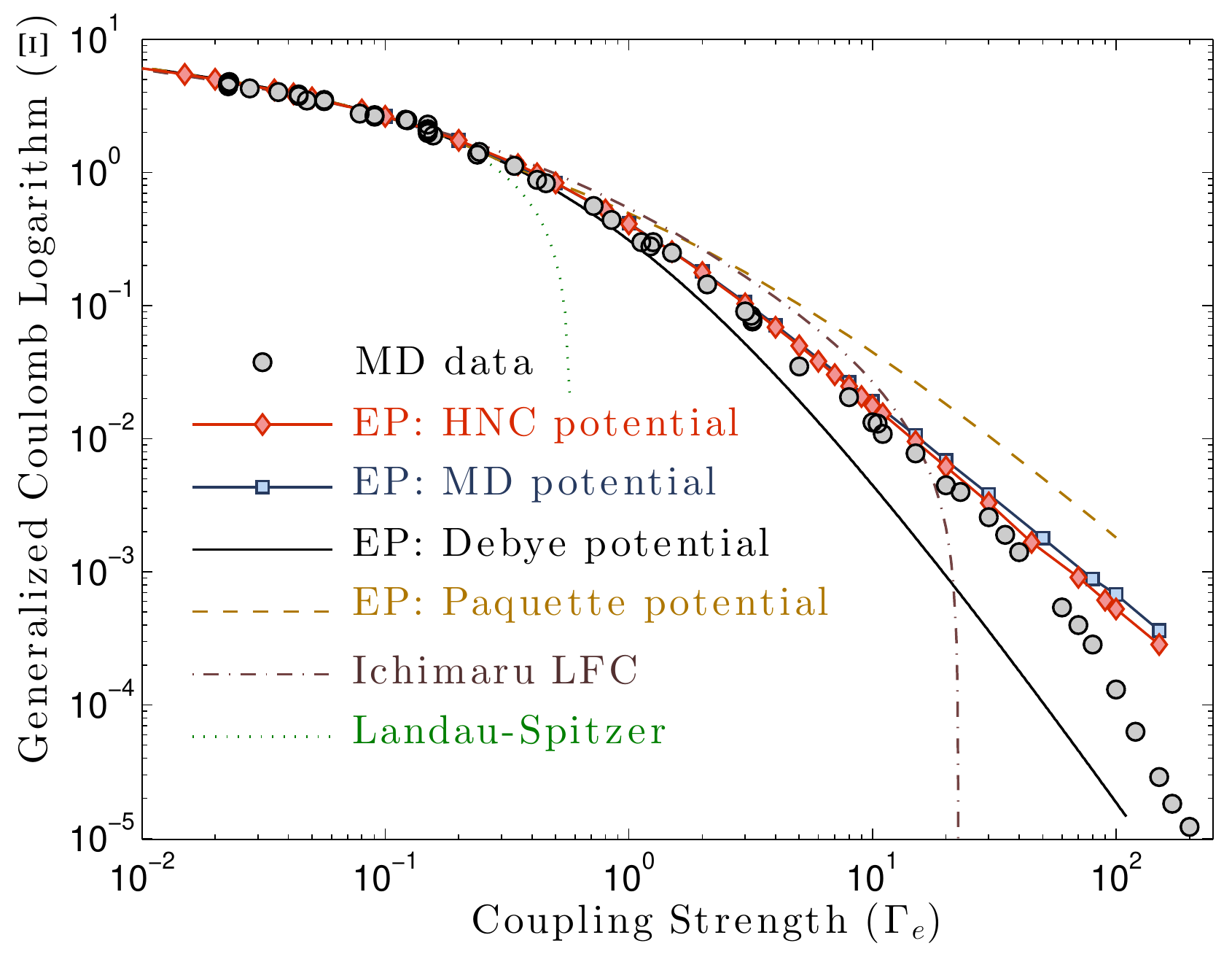}%
\caption{Generalized Coulomb logarithm inferred from MD simulations of positron-ion temperature relaxation (circles). Theoretical predictions are shown for a variety of effective potentials: obtained from HNC (diamonds), obtained from MD simulations (squares), a Debye screened potential (solid line), and a modified Debye screened potential from Paquette \textit{et.\ al.\ } \cite{paqu:86} (dashed). Ichimaru's local field correction theory (dash-dotted line) and the conventional Landau-Spitzer plasma theory (dotted line) are also shown. }
\label{fg:t_r}
\end{figure}

Figure~\ref{fg:t_r} shows the lowest-order generalized Coulomb logarithm obtained from MD simulations in comparison to a compilation of theoretical models. For the MD simulations, this was inferred from the temperature relaxation rate between protons and positively charged electrons (positrons). This was done by fitting the temperature relaxation rate using $dT_e/dt = 2Q^{e-i}/2n_e$ where $\mathcal{Q}^{e-i} = -3m_{ei}n_e \nu_{ei} (T_e-T_i)/m_i$ is the energy exchange density and $\nu_{ei}$ the reference collision frequency from Eq.~(\ref{eq:nu}); see \cite{dimo:08} for details on the simulations. 

Some of the theoretical curves in the figure were obtained using different effective potentials in the theory from Sec.~\ref{sec:ep}. Again, the HNC approximation is shown to provide similarly accurate results as were seen for the other transport coefficients (diamonds). Results obtained using an effective potential extracted directly from the MD simulations, by applying $\phi/(k_BT) = -\ln [g(r)]$ (squares), show that inaccuracies of the HNC approximation for $g(r)$ lead to negligible corrections to the generalized Coulomb logarithm. This is shown to demonstrate that the difference between effective potential theory and MD results is not due to inadequacies of the HNC approximation. The solid line shows results using the screened Coulomb potential as the effective potential: $\phi = \Gamma/(r/a) \exp(-r/\lambda_D)$ where $\lambda_D = \sqrt{k_BT/(4\pi ne^2)}$. This provides a convergent kinetic theory, but it does not capture correlation effects that become essential for $\Gamma \gtrsim 1$. This approach has been applied by a number of previous authors (see \cite{paqu:86,baal:12} and references therein). A model proposed by Paquette \emph{et al} \cite{paqu:86} is also shown (dashed line). This is also a screened Coulomb potential, but where the screening length is $\lambda = \max \lbrace \lambda_D , a \rbrace$. This modified screening length is physically motivated by the fact that the interaction range is characterized by the interparticle spacing, rather than the Debye length, at strong coupling. However, a more accurate approximation is obtained using the HNC potential that includes correlation affects in addition to screening. 

The generalized Lenard-Balescu theory proposed by Ichimaru~\cite{ichi:92,tana:86} is also shown for comparison (dash-dotted line). This was computed from
\begin{equation}
\Xi_{I}= \frac{2}{\sqrt{\pi}} \int_0^\infty dk \frac{[1-G(k)]}{k} \int_0^\infty dz \frac{e^{-z^2}}{|\hat{\varepsilon}(k, k v_Tz)|^2}   \label{eq:ichi}
\end{equation}
where $v_T = \sqrt{k_BT/m}$. Here $\hat{\varepsilon}(k, kv_T z) = 1 - 3\Gamma [1 - G(k)] Z^\prime ( z/\sqrt{2})/(ka)^2$ is the equilibrium dielectric response, $[1-G(k)]$ is the local field correction and $Z$ the plasma dispersion function. Here, the local field correction was calculated using HNC with Ichmaru's bridge function~\cite{ichi:92,tana:86}, which gives access to the direct correlation function $c(k)$ and provides $G(k)= 1 + \frac{k_B T}{v(k)} c(k)$. The figure shows that this approach also provides some extension of traditional plasma theory into the strong coupling regime. However, the Coulomb logarithm becomes negative at a value $\Gamma \simeq 22.4$ for this HNC model (negative values occur for $\Gamma \gtrsim 12$ when using the standard HNC model without a bridge function). It is unknown if this approach would extend significantly farther in coupling strength if a more accurate local field correction could be calculated. Local field correction approaches, such as Eq.~(\ref{eq:ichi}), model strong coupling physics by modifying the density response, rather than the interaction potential.

\section{Summary}

The effective potential theory is a physically motivated approximation for extending traditional plasma transport theories into the strongly coupled regime. It is similar in character to local field correction theories, which model strong coupling physics by modifying a response function describing the interaction of an individual particle with the rest of the plasma. However, local field correction theories generalize the Lenard-Balescu equation through a modified density response~\cite{dali:09,bene:12}, whereas the effective potential theory generalizes the Boltzmann equation through a modified interaction potential. Previous work has shown that the effective potential approach provides an accurate extension for the OCP, as well as ion velocity relaxation in an ultracold neutral plasma~\cite{bann:12,baal:13,baal:14}. Recent work has also applied the effective potential theory to binary ionic mixtures~\cite{haxh:14,bezn:14}, and a similar concept to stopping power~\cite{grab:13}. The present work has extended these results to Yukawa OCP systems, and has compared this approach with previous theories. The results obtained for Yukawa OCP were similar to the classical OCP. 

\begin{acknowledgement}
  The work of S.D.B.\ was supported in part by the University of Iowa. The work of J.D.\ and K.\O.R.\ was supported by the DOE Office of Fusion Sciences.
\end{acknowledgement}

%
%

\end{document}